 \documentstyle[12pt]{article}  

\def\nn{\nonumber} \def\bd{\begin{document}} \def\ed{\end{document}}
\def\ds{\documentstyle} \let\fr=\frac \let\bl=\bigl \let\br=\bigr
\let\Br=\Bigr \let\Bl=\Bigl 
\let\bm=\bibitem
\let\na=\nabla
\let\pa=\partial \let\ov=\overline 
\def\ba{\begin{array}}
\def\ea{\end{array}}
\def\nn{\nonumber}
\newcommand{\ho}[1]{$\, ^{#1}$}
\newcommand{\hoch}[1]{$\, ^{#1}$}
\newcommand{\bea}{\begin{eqnarray}} 
\newcommand{\eea}{\end{eqnarray}} 
\newcommand{\ra}{\rightarrow}
\newcommand{\lra}{\longrightarrow}
\newcommand{\Lra}{\Leftrightarrow}
\newcommand{\ap}{\alpha^\prime}
\newcommand{\bp}{\beta^\prime}
\newcommand{\tr}{{\rm tr} }
\newcommand{\Tr}{{\rm Tr} } 
\newcommand{\NP}{Nucl. Phys. }
\newcommand{\tamphys}{\it Center for Theoretical Physics,
Physics Department, Texas A \& M University, College Station, Texas
77843} 
\newcommand{\auth}{J. X. Lu
\footnote{Research supported by NSF Grant PHY-99411543.}}
\def\ct#1{\cite{#1}}
\def\req#1{(\ref{#1})}
\def\be{\begin{equation}}
\def\ee{\end{equation}}
\def\tr{\hbox{\rm tr}\,}
\def\Tr{\hbox{\rm Tr}\,}
\def\diag{\hbox{\rm diag}\,}
\def\ap{\a^\prime}
\def\bp{\b^\prime}
\def\za#1{{[#1]}}
\def\eqname#1{\label{#1}}
\def\gapij{{1\over 4 \p\ap}\,\sqrt{-\g}\,\g^{ij}}
\renewcommand{\theequation}{\arabic{section}.\relax \arabic{equation}}
\begin{document}

\hfill{CTP-TAMU-44/97}

\hfill{hep-th/9711014}

\vspace{20pt}

\begin{center}
{ \large {\bf Remarks on M-Theory Coupling Constants\\
and M-Brane Tension Quantizations\\}}

\vspace{36pt}

\auth

\vspace{20pt}

{\tamphys}
\vspace{44pt}

%%%%%%%%%%%%%%%%%%%%%%%%%%%%%%%%%%%%%%%%%%%%%%%%%%%%
\makeatletter
\@addtoreset{equation}{section} 
\makeatother
\renewcommand{\theequation}{\thesection.\arabic{equation}}
\addtolength{\baselineskip}{0.3\baselineskip}
%%%%%%%%%%%%%%%%%%%%%%%%%%%%%%%%%%%%%%%%%%%%%%%%%%%%   

\underline{ABSTRACT}
\vspace{8pt}
\end{center}
In the absence of a complete M-theory, 
we gather certain quantum aspects of this theory,
namely,  M-2 and M-5 brane duality and their tension quantization
rule $2\kappa^2 T_2 T_5 = 2 \pi n$, the M-2 brane tension quantization 
$T_2 =\bigg({(2\pi)^2/2\kappa^2 m}\bigg)^{1/3}$, supersymmetry, perturbative
gauge and gravitational anomaly cancellations, and the half-integral
quantization of [$G^W/2\pi$], and study the consistency among these quantum
effects. We find: (1) The complete determination of Ho\u{r}ava-Witten's
 $\eta = \lambda^6 /\kappa^4$ for M-theory on 
$R^{10} \times S^1/ Z_2$ requires not only the cancellation of 
M-theory gauge anomaly but also that of the gravitational anomaly,  
the quantization of M-2 brane tension, and the recently recognized
half-integral quantization of $[G^W / 2 \pi]$. (2)
A well-defined quantum M-theory necessarily requires the presence of both M-2
and M-5 branes and allows only $n = 1$ and $m =1$ for
 the respectively quantized M-2 and M-5 brane tensions. Implications of the 
above along with other related issues are discussed.                           
\vfill \footnoterule
{\footnotesize \vskip -12pt }
\baselineskip=24pt

\pagebreak
\setcounter{footnote}{0}
\setcounter{page}{1}
\section{Introduction}
 Ho\u{r}ava and Witten \cite{horwone,horwtwo} recently proposed that the
strongly
coupled ten dimensional $E_8 \times E_8$ heterotic string  is described by 
 M-theory on $M^{11} = R^{10} \times S^1 /Z_2$. Cancellation of the
gravitational anomaly appearing on
the boundary of space time in the M-theory requires the introduction of 
gauge fields with gauge group a copy of $E_8$ on each boundary component. 
The gauge fields enter via ten dimensional vector
multiplets that propagate on the boundary of space time. This picture
immediately implies that there must exist a supersymmetric coupling of
ten-dimensional vector multiplets on the boundary of the above eleven-manifold
to the eleven-dimensional supergravity multiplet propagating on the bulk.  
In the so-called ``upstairs" approach (We will discuss the ``upstairs" and
``downstairs" approaches later in this section), this has been achieved to 
the lowest 
order in supersymmetry in \cite{horwtwo}
through modifying the Bianchi identity $d G = 0$
for the four-form field
strength $G$\footnote{We will use Ho\u{r}ava
and Witten's notation throughout unless stated otherwise. In their notation, 
$G_{IJKL} = \partial_I C_{JKL} \pm 23$ terms, $C = C_{IJK} dx^I dx^J dx^K$, 
and $dG = {1\over 4!} G_{IJKL} dx^I d x^J dx^K dx^L$. 
The comparison with the notation of
\cite{dufkl} is as follows: $C^{DKL} = 6 \sqrt{2} C$ and 
$G^{DKL} = \sqrt{2} G$.
Our indices I, J, K, L run from 1 to 11 while A, B, C, D run from 1 to 10.}
to  
\begin{equation}
d G = - {\kappa^2 \over \sqrt{2} \lambda^2} \delta (x^{11}) d x^{11} {\rm tr}
F^2 \label{keyone},
\end{equation}
where $\kappa$ is the ${\rm D} = 11$ gravitational coupling constant while
 $\lambda$ is the ${\rm E}_8$ gauge coupling
constant. Based on the known form of the ten dimensional anomalies, 
Ho\u{r}ava and Witten extended Eq.\ (\ref{keyone}) to 
\begin{equation}
d G = {\kappa^2 \over \sqrt{2} \lambda^2} \delta(x^{11}) dx^{11} \hat
{I_4}\label{keytwo},
\end{equation}
where 
\begin{equation}
\hat{I_4} = {1\over 2} {\rm tr} R^2 - {\rm tr} F^2\label{defIfour},
\end{equation}
with tr the trace in the fundamental representation of the
corresponding group (for ${\rm E}_8$, it is defined as 
${\rm Tr} F^2 = 30~ {\rm tr} F^2$ with Tr the trace in the adjoint
representation of ${\rm E}_8$).

Modifying $dG = 0$ to the form of Eq.\ (\ref{keytwo}) implies that the
three-form potential $C$ is in general transformed under a gauge or a local
Lorentz transformation (This is familiar in coupling  $N = 1$
supergravity to super Yang-Mills theory in ten dimensions). The ``Chern-Simons"
interactions $\int C\wedge G\wedge G$ present in the eleven-dimensional
supergravity is therefore not invariant under either of these transformations, 
therefore the presence of anomalies. With the correction 
$\int C\wedge X_8$ to the eleven-dimensional
supergravity action either from membrane-fivebrane
duality based on M-5 brane worldvolume one loop anomalies \cite{duflm} or from
one-loop calculation of Type IIA superstring\cite{vafw}, both the above gauge
and gravitational anomalies can potentially be cancelled. In the above
\begin{equation}
X_8 = - {1\over 8} {\rm tr} R^4 + {1\over 32} ({\rm tr} R^2)^2.\label{defx8}
\end{equation}
For a special
solution of G from Eq.\ (\ref{keytwo}), this is indeed true as shown in 
\cite{horwtwo,deaone} even though numerical errors occurred which has been 
corrected recently in \cite{con}. Requiring anomaly-freedom also fixes the ratio 
of $\lambda^6 /\kappa^4$, i.e., determining the gauge coupling constant in terms of
the gravitational constant.   

Not long after his above work with Ho\u{r}ava, Witten\footnote{$G^W$ is the 
one used by Witten in his paper \cite{wit}.
Its relation to the present $G$  and to  the $G^{DKL}$ given in footnote 1 is:
$G^W \equiv T_2 G^{DKL}  = \sqrt{2}~ T_2 G$.} \cite{wit} pointed out 
that  $G^W / 2 \pi$
should have in general a half-integral period rather than an integral one 
as previously used in establishing the quantization of the M-2 brane tension 
in terms of the gravitational constant 
$\kappa$ as 
\begin{equation}
T_2 = \bigg({(2\pi)^2 \over 2 \kappa^2 m}\bigg)^{1/3}\quad (m = {\rm integer}).
\label{twot}
\end{equation} 
 This is based on the 
observation that there is a sign ambiguity for 
the fermion path integral for fermions on the membrane worldvolume. This
potential inconsistency in defining the sign of the fermion path integral is 
fortunately correlated to the ``Chern-Simons" factor coming from the
coupling of the membrane worldvolume to the three-form potential $C$. A
well-defined membrane path integral can be obtained if $G^W / 2\pi$ has a
half-integral period in general. For the aforementioned 
 special solution\footnote{The final expression for $G^W / 2 \pi$ is correct in
spite of some numerical errors in \cite{horwtwo}.}
of $G$ 
from Eq.\ (\ref{keytwo}), it has been shown in \cite{wit} that $G^W/ 2\pi$ 
has indeed a half-integral period in general.

In this article, we intend to ask ourselves what we can learn, on a general
ground, from the requirement of anomaly-freedom for M-theory on ${\rm
M}^{11} = {\rm R}^{10} \times {S^1 /Z_2}$ and that of a well-defined
membrane path integral, i.e., $G^W / 2\pi$ has a half-integral period in
general.
Contrary to what has been claimed in \cite{horwtwo,deaone} and 
recently in \cite{con},
 the cancellation of pure gauge anomaly or pure gravitational anomaly or 
both anomalies cannot determine  $\eta = \lambda^6 / \kappa^4$ uniquely. 
To completely determine this $\eta$, we need in addition the following: (1)
M-2 and M-5 brane duality and the associated quantization 
rule\cite{duflm}\footnote{In \cite{duflm}, $n$ was imposed to be
a non-negative integer from the fact that a p-brane tension is the measure of 
the energy per unit p-brane volume and it should be non-negative. Recently, 
these tensions are also allowed to be negative purely from the 
viewpoint of classical solutions from supergravity theories. Past experiences
tell us that these extended objects with negative tensions cannot be stable
quantum mechanically. We therefore insist that $n$ be non-negative throughout
this paper since we are considering the quantum effects of M-theory. Hence the
integer $m$ in Eq.\ (\ref{twot}) should also be non-negative. But as we will
see in the next section, this will follow automatically once $n$ is
non-negative. If the non-negative condition for $n$ is dropped, the solution of
$\alpha = 1$, $m = -1$ and $n = -1$ is also allowed in addition to the one
obtained in the next section.} 
\begin{equation}
2 \kappa^2 T_2 T_5 = 2\pi n\quad ({\rm n~ =~
integer}),\label{twofivet}    
\end{equation}
 (2) $G^W / 2 \pi$ has a
half-integral period in general,  (3) M-2 brane tension $T_2$ is 
quantized according to Eq.\ (\ref{twot}).
There exist also alternatives of the above which we will discuss in the next
section. The other lesson of this investigation is that a well-defined quantum
M-theory requires the presence of both M-2 and M-5 branes and 
 allows only M-2 brane tension $T_2 =  \bigg({(2\pi)^2 \over 2
\kappa^2}\bigg)^{1/3}$ and M-5 brane tension $T_5 =  \bigg({ 2\pi \over (2
\kappa^2)^2}\bigg)^{1/3}$, i.e., only $m = 1$ and $n =1$ are allowed 
in Eqs.\ ({\ref{twot}) and \ (\ref{twofivet}), respectively.

Two things remain to be discussed before we move on to the next section. 
One is about the M-2 brane tension quantization of Eq.\ (\ref{twot}). 
The other is about clarifying the role of the ``upstairs" and 
``downstairs" approaches. 

  The knowledge we learned from perturbative
string theories about the relations among various coupling constants indicates
that there might be only one independent constant in ${\rm D} = 11$ M-theory in
the absence of a dilaton. This implies that there exists a relation
between the M-2 brane tension $T_2$ and the gravitational constant $\kappa$ 
(similarly, a relation between $T_5$ and $\kappa$).
 For a special choice of a twelve manifold as ${\rm
Q} = D^4 \times D^4 \times D^4$, this relation has been established as 
given by Eq.\ (\ref{twot}) in \cite{duflm,deatwo}. 
For this special choice of manifold, the $m$ in Eq.\ (\ref{twot}) remains 
still as an integer even with the 
recent work of Witten \cite{wit} that $G^W /2\pi$ has a half-integral period in
general
and  $\int_Q G\wedge G\wedge G$ is half-integral in general. One cannot
derive this formula for a general twelve manifold $Q$. The arguments presented
above nevertheless support  that Eq.\ (\ref{twot}) with $m = {\rm integer}$
 should be true in general. 

In the ``upstairs" approach of Ho\u{r}ava and Witten \cite{horwtwo},
the bulk action
of M-theory on ${\rm R}^{10} \times S^1 / Z_2$ is given by
\begin{equation}
S_M^U = - {1\over 2 \kappa^2} \int_{M^{11}_U} d^{11} x \sqrt{-g_M} (R +
\cdots),\label{hw}
\end{equation}
where $M^{11}_U = R^{10} \times S^1$, all fields are $Z_2$ symmetric, $Z_2$
is generated by $x^{11} \mapsto - x^{11}$ and ``$\cdots$" are the terms 
involving three-form field and fermions. In the above, the $\kappa^2$ is the one used  
 in the usual eleven dimensional supergravity which
is believed to be the low-energy effective description of M-theory 
in ${\rm D} = 11$. 
Recently, Conrad \cite{con} claimed that the $\kappa^2$ appearing in the 
above ``upstairs" action should be replaced by $2 \kappa^2$. By this, 
Conrad  obtained the following 
``downstairs" action following Ho\u{r}ava and Witten\cite{horwtwo}
\begin{equation}
S^D_M = - {1 \over 2 \kappa^2} \int_{M^{11}_D} d^{11} x \sqrt{- g_M} (R +
\cdots),\label{conrad}
\end{equation}
where $M^{11}_D = {\rm R}^{10} \times S^1/Z_2 ={\rm R}^{10} \times I$.

In what follows, I will argue that Ho\u{r}ava and Witten's original ``upstairs"
action given by Eq.\ (\ref{hw})
is with the correct unit but their ``downstairs" action given
in \cite{horwtwo} cannot be the bulk action describing the local physics
observed in the bulk when the radius of the $S^1/Z_2$ is large. 
On the other hand,
Conrad's above proposed ``downstairs" bulk action is with the correct unit
but not his ``upstairs" action.   

Our arguments are based on the following: (1) The ``upstairs" and ``downstairs"
approaches each should give the same results when implemented properly. (2) The 
bulk M-2 and M-5 brane tensions are independent of whether any boundaries exist
arbitrarily far away from a local observer or whether a dimension is compact at 
arbitrary large scales\footnote{This point is borrowed from \cite{con}.}. 
(3) The bulk Lagrangian constructed by the local observer based on local 
symmetries
such as the local supersymmetry should be the same. (4) To
the local observer, the equations of motion  describing, for example, a M-2
brane with a given tension $T_2$ in the bulk background fields, whether they
are derived from the ``upstairs" bulk background field action plus the M-2
brane worldvolume action or from the ``downstairs" correspondent, should be the
same as those from the usual eleven dimensional supergravity action plus a M-2 
brane worldvolume action, e.g., those given by Duff and Stelle\cite{dufs}. 

The above points, especially (4), immediately imply that the correct
``upstairs" bulk action is given by Eq.\ (\ref{hw}) while the correct 
``downstairs" bulk action is given by Eq.\ (\ref{conrad}). 
Otherwise, we would obtain
different equations of motion describing the M-2 brane moving in the bulk
from ``upstairs" and 
``downstairs" approaches since the observer has the same worldvolume action
describing the M-2 brane with a given tension $T_2$ in both cases. 

In the above sense, the ``downstairs" bulk action can be effectively 
identified with 
either $x^{11} > 0$ or $x^{11} < 0$ component of the ``upstairs" bulk action 
with the $Z_2$ symmetry imposed. However, as the radius of $S^1/Z_2$
approaches zero, i.e., taking the
weakly coupled limit of the heterotic string, for which the sense of bulk is
diminishing, two copies of the ``downstairs" bulk action Eq.\ (\ref{conrad}) 
have to be used such that the correct ${\rm N} = 1~~ {\rm
D} = 10$ supergravity can be obtained, i.e., in this limit, Ho\u{r}ava and 
Witten's ``downstairs" action is with the correct normalization.
Therefore, in the ``downstairs" approach, studies of the aforementioned bulk 
properties as well as the perturbative gauge and gravitational anomalies of 
the M-theory need Conrad's ``downstairs" action while in obtaining the correct 
weakly coupled low energy effective action of the heterotic string, we need 
Ho\u{r}ava and Witten's ``downstairs" action.
An independent check will be provided in section 3, based on a recent work 
\cite{bram}, that Conrad's  ``downstairs" action is needed in order to cancel
both gauge and gravitational anomalies.                      
We will never encounter the above complications if the ``upstairs" approach is
employed. As we will see in the next section, there is actually a subtle
difference between these two approaches. 

The above discussion clearly demonstrates the point of Ho\u{r}ava and Witten
\cite{horwtwo} that even though the ``upstairs" approach does not manifest the
feature of M-theory on $R^{11} \times S^1/Z_2$, it is indeed convenient for
calculation. On the other hand, the ``downstairs" approach is just the other
way around.   Carrying out the calculations properly in the
``downstairs" approach is explained well in \cite{horwtwo,deaone,con}. In the
``upstairs" approach, we always begin with the action as if it is on 
$R^{10}\times S^1$ or simply on $R^{11}$. Only at the last 
step in obtaining the results we need, the $Z_2$ symmetry is imposed on the 
fields.

\section{Analysis of Anomalies}
\subsection{``Upstairs" Approach}

We begin with the ``upstairs" approach.
For all the previous studies, a specific solution of Eq.\ (\ref{keytwo}) was 
always chosen  \cite{horwtwo,deaone,con}. 
This equation actually has the following 
general solution  
\begin{equation}
G = 6 d C + \alpha {\kappa^2 \over \sqrt{2} \lambda^2} 
\delta(x^{11}) dx^{11} Q_3 + 
{(1 + \alpha) \over 2} {\kappa^2 \over \sqrt{2} \lambda^2} \epsilon (x^{11}) 
\hat{I_4} \label{gsolution},
\end{equation}
 where $\alpha$ is an as yet undetermined dimensionless constant, 
$Q_3 = {1\over 2} \omega_{3L} - \omega_{3Y}$ and $\epsilon (x^{11})$ is a step
function such that $\epsilon (x^{11}) = - \epsilon (- x^{11}) = 1$ if $x^{11} >
0$ and $d\epsilon (x^{11}) /d x^{11} = 2 \delta (x^{11})$. This
general solution was also given recently in \cite{dudm}.

In terms of components, we have
\begin{equation}
G_{11 ABC} = 4! \partial_{[11} C_{ABC]} + \alpha {\kappa^2 \over \sqrt {2}
\lambda^2} \delta(x^{11}) Q_{3 ABC}, \label{g11abc}
\end{equation}
and
\begin{equation}
G_{ABCD} = 4! \partial_{[A} C_{BCD]} + {4! (1 + \alpha) \over 8} {\kappa^2
\over \sqrt {2} \lambda^2} \epsilon (x^{11}) \bigg[ {1\over 2} 
{\rm tr} R_{[AB}R_{CD]} - {\rm tr} F_{[AB}F_{CD]}\bigg]\label{gabcd},
\end{equation}
where $C_{ABC} = 0$ at $x^{11} = 0$.

For comparison with Ho\u{r}ava and Witten's result of the pure gauge anomaly 
cancellation \cite{horwtwo}, we consider the gauge anomaly first. 
Under a gauge transformation
$\delta A = - D \epsilon (x)$, from $\delta G = 0$ we deduce 
\begin{equation}
\delta C = - {\alpha \over 3!} {\kappa^2 \over \sqrt{2} \lambda^2} \delta
(x^{11}) dx^{11} Q_{2Y}^1\label{deltac},
\end{equation}
where $Q_{2Y}^1 = - {\rm tr}\epsilon F$. In components,
\begin{equation}
\delta C_{11 AB} = {\alpha \over 3!} {\kappa^2 \over \sqrt{2} \lambda^2} 
\delta (x^{11}) {\rm tr} \epsilon F_{AB}\label{deltac11ab},
\end{equation}
and $\delta C_{ABC} = 0$. Under this gauge transformation, the ``Chern-Simons"
interaction term
\begin{equation}
W = - {\sqrt{2} \over 3456 \kappa^2} \int_{M^{11}_U} d^{11} x \epsilon^{M_1
M_2\cdots M_{11}} C_{M_1 M_2 M_3} G_{M_4 \cdots M_7} G_{M_8 \cdots M_{11}}
\label{chernsimons},
\end{equation}
 in the classical action of ${\rm D} = 11$ supergravity is not invariant 
but transforms as 
\begin{equation}
\delta W = - {\alpha (1 + \alpha)^2 \over 1536} {\kappa^4 \over \lambda^6} 
\int_{M^{10}} d^{10} x \epsilon^{A_1 A_2 \cdots A_{10}} {\rm tr} \epsilon 
F_{A_1 A_2} {\rm tr}F_{A_3 A_4} F_{A_5 A_6}
{\rm tr} F_{A_7 A_8} F_{A_9 A_{10}}\label{wanomaly}.
\end{equation}
To cure such a gauge non-invariance of the classical theory, we have to 
appeal to quantum anomalies. In the present case that the gauge group is 
$E_8$ and the Majorana-Weyl fermions are in the adjoint representation, 
the anomalous variation of the effective action $\Gamma$ for the 
ten-dimensional fermions 
is \footnote{There seems short of a factor ${1\over 6}$ in Eq. (3.5) of
\cite{horwtwo} for $\delta \Gamma$ (where a ${1\over 2}$ factor is indeed 
included for Majorana-Weyl fermions). This can be 
easily checked if one uses the anomalous 12-form, which should be halved for
Majorana-Weyl spinors, for gauge fields 
from Green-Schwarz-Witten \cite{gresw}. Using the standard procedure, we obtain
Eq.\ (\ref{gaugeanomaly}). This was also pointed out in \cite{con}.}
\begin{equation}
\delta \Gamma = {1 \over 2} {1 \over (4 \pi)^5}{1 \over 6!} 
\int_{M^{10}} d^{10} x 
\epsilon^{A_1 A_2 \cdots A_{10}} {\rm Tr}
(\epsilon F_{A_1 A_2} F_{A_3 A_4} \cdots
F_{A_9 A_{10}})\label{gaugeanomaly},
\end{equation}
where Tr is the trace in the adjoint representation of gauge group $E_8$. 
As in \cite{horwtwo},
for $E_8$ we have the identity ${\rm Tr}W^6 = ({\rm Tr}W^2)^3/7200$ 
(and likewise, 
${\rm Tr}\epsilon F^5 = {\rm Tr}\epsilon F (Tr F^2)^2 / 7200$) and the relation 
${\rm tr} W^2 = {\rm Tr} W^2 /30$. Then we have 
${\rm Tr} \epsilon F^5 = (15/4) {\rm tr}\epsilon F ({\rm tr} F^2)^2$. 
With this, 
Eq.\ (\ref{gaugeanomaly}) can be rewritten as
\begin{equation}
\delta \Gamma = {1 \over 16 (4 \pi)^5 4!} \int_{M^{10}} d^{10} x
 \epsilon^{A_1 A_2 \cdots A_{10}} 
{\rm tr}\epsilon F_{A_1 A_2} {\rm tr} F_{A_3 A_4} F_{A_5 A_6} 
{\rm tr} F_{A_7 A_8} F_{A_9 A_{10}}\label{gaugea}.
\end{equation}
Setting $\delta W + \delta \Gamma = 0$, we have
\begin{equation}
\eta = {\lambda^6 \over \kappa^4} = {\alpha (1 + \alpha)^2 \over 4} (4 \pi)^5.
\end{equation}

Unlike in \cite{horwtwo}, we can determine the $\eta$ only up to an as yet
undetermined  constant $\alpha$ if only the gauge anomaly cancellation is 
imposed.
That Ho\u{r}ava and Witten can determine the $\eta$ uniquely at this 
stage is because they chose a specific solution, i.e., setting $\alpha = 1$ 
from the outset in Eq.\ (\ref{gsolution}), for  $G_{11 ABC}$ and $G_{ABCD}$
\footnote{Actually, for 
$\alpha = 1$, the present $G_{11ABC}$ agrees with theirs but the present 
$G_{ABCD}$ are
twice theirs. This short of factor 2 fortunately gives the correct expression
for $G^W /2 \pi$ even though their $\eta$ is short of a factor 8.}. 
In the following, we will show that to determine the 
$\eta$ completely, more conditions are needed as discussed
in the introduction. 

We now consider both gauge and gravitational anomaly cancellations\footnote{This
has been considered in \cite{deaone} for the case of $\alpha = 1$ in the
``downstairs" approach. 
However, the variation of the quantum
effective action $\delta \Gamma$ used there is for Weyl fermions but not for
Majorana-Weyl fermions. In other words, a factor 2 is overcounted in the 
$\delta \Gamma$ there. If the correct $\delta \Gamma$ for Majorana-Weyl spinors
is used, the gravitational anomaly discussed in \cite{deaone} would not be
cancelled if Ho\u{r}ava and Witten's $\eta = 2^7 \pi^5$ is used.}. The
``Chern-Simons" interaction of Eq.\ (\ref{chernsimons}) can be re-expressed as 
\begin{equation}
W = - {\sqrt{2} \over \kappa^2} \int_{M^{11}_U} C \wedge G \wedge G\label{neww}.
\end{equation}
In order to determine the variation of W 
under both a gauge transformation $\delta A = - D \epsilon$ and a local Lorentz 
variation $\delta \omega = - D \Theta$ with $\omega$ the spin connection, we
have to determine the variation of the 3-form $C$ first. It can be deduced from 
$\delta G = 0$ as 
\begin{equation}
\delta C = {\alpha \over 3!} {\kappa^2 \over \sqrt{2} \lambda^2} \delta(x^{11}) 
dx^{11} Q_2^1,
\end{equation}
where 
\begin{equation}
Q_2^1 = - \bigg({1 \over 2} {\rm tr} \Theta R - {\rm tr}\epsilon F\bigg)
\label{parameter}.
\end{equation}
Then we have
\begin{equation}
\delta W = - {\alpha (1 + \alpha)^2 \over 12}{\kappa^4 \over \lambda^6}
\int_{M^{10}} Q_2^1 \wedge {{\hat I_4}^2 \over 4}.
\end{equation}
There is an additional Green-Schwarz term which appears in M-theory whose
existence can be inferred either from ${\rm D} = 11$ membrane-fivebrane duality
and the world-volume anomaly cancellation of M-5
fivebrane \cite{duflm} or from a one-loop calculation of Type IIA
superstrings\cite{vafw}. It is in general
\begin{equation}
W_5 = {1 \over 2 \sqrt{2} (2\pi)^4} {T_2 \over n} \int_{M^{11}_U} C\wedge
X_8\label{fanomaly},
\end{equation}
where $n$ is the non-negative integer appearing in Eq.\ (\ref{twofivet}) 
 and $T_2$ is the M-2 brane tension which is quantized according to 
Eq.\ (\ref{twot}), and the
8-form $X_8$ is given in Eq.\ (\ref{defx8}).
Under the above gauge and local Lorentz variations,
\begin{equation}
\delta W_5 = {1 \over 24 (2\pi)^4} {\alpha \over n}
\bigg({(2\pi)^2 \over 2 m} {\kappa^4
\over \lambda^6}\bigg)^{1/3} \int_{M^{10}} Q_2^1 \wedge X_8 \label{delwfive}.
\end{equation}

Since $W +  W_5$ is not invariant under either gauge or local Lorentz
variation, we have to appeal to quantum anomalies. The variation of quantum
effective action $\Gamma$ for ten-dimensional Majorana-Weyl fermions in the
present case is
\begin{equation}
\delta \Gamma = - {1\over 2} {1 \over 48 (2\pi)^5} \int_{M^{10}} Q_2^1 \wedge
\bigg( - {{\hat I}_4^2 \over 4} + X_8 \bigg)\label{quantuma},
\end{equation}
which is the half of the $\delta \Gamma$ used in \cite{deaone}.

Cancellations of both gauge and gravitational anomalies imply $\delta W +
\delta W_5 + \delta \Gamma = 0$. This gives
\begin{equation}
{\lambda^6 \over \kappa^4} = {\alpha (1 + \alpha)^2 \over 4} (4\pi)^5,
\label{eta}
\end{equation}
and 
\begin{equation} 
{\alpha \over n}
\bigg( {(2\pi)^2 \over 2 m } {\kappa^4 \over \lambda^6} \bigg)^{1/3}  =
{1 \over 8 \pi}.\label{mnsign}
\end{equation}
Solving the above two equations gives
\begin{equation}
\alpha = {\sqrt{m n^3} \over 2 - \sqrt{m n^3}}\label{alpha}.
\end{equation}
From Eq.\ (\ref{eta}), we have $\alpha > 0$. Applying this to the above
equation, we have 
\begin{equation}
0 < m n^3 < 4,
\end{equation}
which has the following solutions: $m = 1, n =1$; $m = 2, n = 1$; and $m = 3, n
= 1$ since both $n$ and $m$ are non-negative integers (we could include the 
two cases corresponding to $\alpha = 0$ and $\alpha = \infty$ in the above, one
corresponding to zero gauge coupling while the other to infinity large gauge
coupling). The corresponding $\alpha$ are 
$1$, $ \sqrt{2}/ (2 - \sqrt{2})$ and $\sqrt{3}/(2 - \sqrt{3})$, respectively.

If we simply stop here,  we must conclude that $\eta = {\lambda^6 \over
\kappa^4}$  is  quantized according to each pair of  ($m$,  $n$) given in the
above. However, 
the other condition, namely the half-integral period of $G^W /2
\pi$, will pick the pair ($m = 1$, $n =1$), therefore, uniquely determining
 the $\eta$. 

From Eq.\ (\ref{gabcd}), we have $G$
on  the ($x^{11} = 0$) component of the boundary as
\begin{equation}
G\mid_N = {1 + \alpha \over 2} {\kappa^2 \over \sqrt{2} \lambda^2}
\hat{I}_4.\label{gatbn}
\end{equation}
Therefore, from footnote 2, we have
\begin{eqnarray}
{G^W \over 2\pi} &=& \sqrt{2}~ T_2~ {G\over 2\pi},\nonumber\\
                 &=& \bigg({1 + \alpha\over 2\alpha m}\bigg)^{1/3} 
{1 \over 16 \pi^2} \bigg({1\over 2} {\rm tr} R^2 - {\rm tr} F^2\bigg),
\label{halfp}
\end{eqnarray}
where Eq.\ (\ref{twot}) has been used.
According to what has been discussed by Witten \cite{wit}, a well-defined
membrane path integral must require 
\begin{equation}
\bigg({1 + \alpha \over 2 \alpha m}\bigg)^{1/3} = l\quad (l = {\rm
odd~integer}),
\end{equation}
such that $G^W/2\pi$ has in general a half-integral period. So we have
\begin{equation}
1 + \alpha = 2 \alpha m l^3.\label{case3}
\end{equation}
Combining this equation with Eq.\ (\ref{alpha}), we have 
\begin{equation}
l^3 (mn)^{3/2} = 1,\label{lmne}
\end{equation}
which has the following unique solution 
\begin{equation}
l = 1, \qquad m = 1,\qquad n = 1, \qquad{\rm and}\qquad \alpha = 1,\label{sol1}
\end{equation}
since both $m$ and $n$ are non-negative integers\footnote{As one can see, the
above conclusion is crucially based on the assumption that $m$ is an integer. 
If we relax $m$ to be a real non-negative number, some conclusions can still 
be drawn. $m$ must be bounded below $4 / n^3$ and is given by $1 / n l^2$ with 
$l$ an positive odd integer satisfying $2 l > n$. Then we have 
$\alpha = n / (2 l -n)$ which implies a quantized  $\eta$ given according to
Eq.\ (\ref{eta}).}. It is also clear from Eq.\ (\ref{lmne}) that $l$ cannot 
be an even integer. This implies that even if we naively assume an integral
period of $G^W /2 \pi$ from the outset, the above process will force us to
conclude a general half-integral period for $G^W/ 2\pi$.

The above uniquely determines 
\begin{equation}
\eta = {\lambda^6 \over \kappa^4} = (4\pi)^5,
\end{equation}
which is eight times that of Ho\u{r}ava and Witten's.
This value of $\eta$ was also obtained recently by Conrad\cite{con}. 
With it, we have the needed expression 
\begin{equation}
{G^W\over 2 \pi} = {1 \over 16 \pi^2} \bigg({1 \over 2} {\rm tr} R^2 - 
{\rm tr} F^2\bigg),
\end{equation}
 for showing that $G^W /2\pi$ has indeed a half-integral period in general.

To summarize, the consistency of quantum M-theory must imply:
(1) Both M-2 and M-5 branes must be present. (2) Only the minimum positive
integers $m = 1$ and $n = 1$ are allowed in Eq.\ (\ref{twot}) and 
Eq.\ (\ref{twofivet}), respectively. This in turn implies that a well-defined
quantum description of M-2 brane may be possible only for a M-2 brane with 
tension $T_2 = \bigg({(2\pi)^2 \over 2 \kappa^2}\bigg)^{1/3}$. So is for a M-5
brane with tension $T_5 = \bigg({2 \pi \over (2 \kappa^2)^2}\bigg)^{1/3}$.
(3) The value of  $\eta = \lambda^6/\kappa^4$ can be determined uniquely to be 
$(4\pi)^5$.

\subsection{``Downstairs" approach}

To demonstrate the usefulness of different approaches, we here repeat the same
process of the previous subsection in the ``downstairs" approach following
the procedure described in \cite{deaone}.

The key for the present approach is Eq.\ (\ref{gatbn}), i.e., on each component 
of the boundary,
\begin{equation}
G\mid_N = {1 + \alpha \over 2} {\kappa^2 \over \sqrt{2} \lambda^2}
\hat{I}_4.\label{rgatbn}
\end{equation}
Under a gauge and a local Lorentz variations described in the previous
subsection, we have standard descent equations
\begin{equation}
\hat{I}_4 = d Q_3, \qquad \delta Q_3 = d Q_2^1,\label{de}
\end{equation}
where $Q_3 = 1/2 \omega_{3L} - \omega_{3L}$ and $Q^1_2$ is given by 
Eq.\ (\ref{parameter}). In the ``downstairs" approach, four-form field strength 
$G$ is always defined as $G = 6 d C$ but with its boundary value given by 
Eq.\ (\ref{rgatbn}). Using the first equation in \ (\ref{de}), we 
therefore have (up to an irrelevant exact form)
\begin{equation}
C\mid_N = {1 + \alpha \over 12} {\kappa^2 \over \sqrt{2} \lambda^2}
Q_3.\label{catbn} 
\end{equation}
It follows using the second equation in \ (\ref{de})
\begin{equation}
\delta C\mid_N = {1 + \alpha \over 12} {\kappa^2 \over \sqrt{2} 
\lambda^2}  d Q^1_2.
\end{equation}
Following \cite{deaone}, we extend this variation to the bulk by writing
\begin{equation}
\delta C = {1 + \alpha \over 12} {\kappa^2 \over \sqrt{2} \lambda^2}
d Q^1_2.\label{vcatb}
\end{equation}
The ``Chern-Simons" interaction in the ``downstairs" version is again given by
Eq.\ (\ref{neww}) but with the replacement of $M^{11}_U$ by $M^{11}_D$.
Therefore we have
\begin{eqnarray}
\delta W &=& - {\sqrt{2} \over \kappa^2} \int_{M^{11}_D}  
{1 + \alpha \over 12} {\kappa^2 \over \sqrt{2} \lambda^2} d Q^1_2 \wedge G
\wedge G,\nonumber\\
&=& - {1\over 3} \bigg({1 + \alpha \over 2}\bigg)^3 {\kappa^4 \over \lambda^6}
\int_{M^{10}} Q^1_2 \wedge {\hat{I}^2_4 \over 4},\label{vw}
\end{eqnarray}
where in reaching the second equality we have used Stokes' theorem, $d G = 0$
and Eq.\ (\ref{rgatbn}). 

Similarly, in the present approach, we have the variation of $W_5$ as
\begin{equation}
\delta W_5 = {1 + \alpha \over 48 n} {1 \over (2\pi)^4} \bigg({(2\pi)^2 
\over 2 m} {\kappa^4 \over \lambda^6}\bigg)^{1/3} \int_{M^{10}} 
Q^1_2 \wedge X_8,\label{dwfive}
\end{equation}
where $T_2 = \bigg({(2\pi)^2 \over 2 \kappa^2 m}\bigg)^{1/3}$ and 
Eq.\ (\ref{vcatb}) have been used.

Again $\delta W + \delta W_5$ does not vanish. But $\delta W + \delta W_5 +
\delta \Gamma = 0$ are possible provided
\begin{equation}
{\lambda^6 \over \kappa^4} = 4 (1 + \alpha)^3 (2\pi)^5,\label{dc1}
\end{equation}
and
\begin{equation}
{1 + \alpha \over n} \bigg({(2\pi)^2 \over 2 m} {\kappa^4 \over
\lambda^6}\bigg)^{1/3} =  {1 \over 4\pi}.\label{dc2}
\end{equation}
In the above, Eq.\ (\ref{quantuma}) for quantum anomalies has been used.
Combining the above two equations, we have
\begin{equation}
n m^{1/3} = 1,\label{nme}
\end{equation}
whose only solution is $n =  1$ and $m =  1$ since both $n$ and $m$ are
non-negative integers\footnote{We can also relax $m$ to be a non-negative real
number. By this, we can determine $m = 1 / n^3$ from Eq.\ (\ref{nme}). So we
have M-2 brane tension $T_2 = n \bigg({(2\pi)^2 \over 2 \kappa^2}\bigg)^{1/3}$ 
 from Eq.\ (\ref{twot}). Applying this to Eq.\ (\ref{twofivet}), we have M-5
brane tension $T_5 = \bigg({2\pi \over 2\kappa^2}\bigg)^{1/3}$, i.e., only the
 minimum M-5 brane tension is allowed.}.  

We have determined $m$ and $n$ completely but not for  $\alpha$. 
One may think that this $\alpha$ can be fixed uniquely 
if we use the condition that $G^W / 2 \pi$ has 
a half-integral period in general as we did in the ``upstairs" approach. 
Unfortunately,  the above solutions
\begin{eqnarray}
&&{\lambda^6 \over \kappa^4} = 4 (1 + \alpha)^3 (2 \pi)^5,\label{ds1}\\
&& n =  1,\qquad m =  1, \label{ds2}
\end{eqnarray}
already imply that 
\begin{eqnarray}
{G^W \over 2 \pi} &=& \sqrt{2}~ T_2~ {G \over 2 \pi},\nonumber\\
                  &=& {1 \over 16 \pi^2} \bigg({1\over 2} {\rm tr} R^2 - 
{\rm tr} F^2 \bigg),
\end{eqnarray}
i.e., $G^W/2 \pi$ has a half-integral period in general.  In other words, in
the ``downstairs" approach, the $G$-value on the boundary can be defined up to
a dimensionless constant.  $G^W / 2\pi$ has always a half-integral period in
general provided no gauge and gravitational anomalies.  

This may make people wonder what happens. However, no contraction exists 
between these
two approaches. One can check that the unique solutions for 
$\alpha = 1$,  $n = 1$ and $m = 1$ obtained in the 
``upstairs" approach continue to be a special case of the present 
solutions. Particularly, if we set $\alpha = 1$ in Eq.\ (\ref{ds1}), we obtain
again $\eta = (4\pi)^5$. 

There actually exists a subtle difference between
these two approaches. In the ``downstairs" approach, under a gauge and a local
Lorentz variations, we deduce the variation for the three-form $C$ all from the
property of $G_{ABCD}$ on the boundary. In the ``upstairs" approach, we have an 
additional information from $G_{11ABC}$. It is just this additional information
that enables us to determine the aforementioned  quantities uniquely.

The fact that a half-integral period of $G^W /2\pi$ is warranted after we 
impose the anomaly-free condition in the ``downstairs" approach may provide
a new way for us to determine uniquely the values for $\alpha$, $n$ and $m$. 
i.e., we identify those conditions obtained from  gauge and gravitational 
anomaly cancellations in both the ``upstairs" and ``downstairs" approaches. 

The ``downstairs" approach already determines $m = 1$ and $n = 1$. Either
substituting $m n^3 = 1$ from Eq.\ (\ref{nme}) in the ``downstairs" approach to 
Eq.\ (\ref{alpha}) in the ``upstairs" approach or identifying
 Eq.\ (\ref{eta}) in the ``upstairs" approach with Eq.\ (\ref{dc1}) in the
``downstairs" approach, we have $\alpha = 1$\footnote{In this new approach, we
can still gain some information if we relax $m$ to be a non-negative real
number. We can determine uniquely $\alpha = 1$, therefore $\eta = (4\pi)^5$,
 and other results given in the previous footnote.}.

\section{Consistency Check}

We begin with a discussion showing that Conrad's ``downstairs" bulk action is 
indeed correct, based on the recent work by Brax and Mourad
\cite{bram}.
Ho\u{r}ava and Witten's work of M-theory on $R^{10} \times S^1/Z_2$
\cite{horwone,horwtwo} nevertheless suggests the existence of an open
supermembrane with each of its two ends lying on each component of the 
boundary of spacetime.  Recently, Brax and Mourad went one step further to 
construct a worldvolume action for  an open supermembrane moving in a flat
spacetime with topological defects on which the membrane can end. This action
is kappa-symmetric and has global spacetime supersymmetry.
To respect the gauge symmetry of the spacetime super three-form potential $C$, 
a spacetime super two-form potential must be introduced whose pullback
contributes a new membrane boundary term, which is absent for a closed
membrane, to the open membrane action. In respect to kappa symmetry, one can
define two field strengthes as (in our notation)
\begin{equation}
G = 6 d C,\qquad H = 6 d B + 6 C.
\end{equation}
The above two field strengthes are not independent but related to each other 
on a topological defect as
\begin{equation}
d H = G\mid_N.
\end{equation}

Consider this open membrane to move in a curved spacetime with a boundary on  
 which super Yang-Mills fields propagate. In order to preserve the
kappa-symmetry and to obtain one of the heterotic strings in the weak coupling
limit, one finds that the unique conclusion is  Ho\u{r}ava and Witten's
\cite{horwone,horwtwo}  but now from
worldvolume rather than spacetime perspective.
At the same time, one finds that the field strength $G$ (It is now the field
strength of the background three-form potential $C$ used in the previous
sections) has to be modified (in our notation) 
on the ($x^{11} = 0$) component of the boundary as
\begin{equation}
d H = G\mid_N = {1\over 8\sqrt{2} \pi T_2} \hat{I}_4,\label{bmkey}
\end{equation}
where $T_2$ is the membrane tension and $\hat{I}_4$ is given by 
Eq.\ (\ref{defIfour}) (for detail, see \cite{bram}).

In the case of Ho\u{r}ava and Witten, $G$-value on
the boundary is essentially determined by the requirement of spacetime 
supersymmetry at the lowest quantum order (the Yang-Mills action is
the quantum correction to the supergravity action)  and the
coefficient in front of $\hat{I}_4$ is in terms of the gauge and gravitational
constants. While in the present case, $G$-value on the boundary is determined by
the requirement of preserving kappa-symmetry at worldvolume one loop level and 
the above coefficient is in terms of the membrane tension. One therefore
expects that the above two $G$-values on the boundary should agree with each
other certainly in a non-trivial way, knowing the fact that 
kappa-symmetry links between spacetime and worldvolume
supersymmetries. 

As pointed out by Brax and Mourad in \cite{bram}, 
the introduction of the two-form potential $B$ on the boundary of spacetime and
its linkage to the two-form potential appearing in the low energy effective
action of the heterotic string in the weak coupling limit must imply, under 
a gauge and a local Lorentz variations described in section 2, 
\begin{equation}
\delta C = 0,\qquad \delta B\mid_N = {1 \over 3!} {1 \over 8 \sqrt{2} \pi T_2}
 Q^1_2,\label{bmv}
\end{equation}
where $Q^1_2$ is given by Eq.\ (\ref{parameter}).

The bulk topological terms in the low energy limit of M-theory on $R^{10} \times
S^1/Z_2$ in the ``downstairs" approach with Ho\u{r}ava and Witten's
 normalization as used by Brax and Mourad in \cite{bram} are
\begin{equation}
S_T = - {2 \sqrt{2} \over  \kappa^2} \int_{M^{11}_D} C \wedge G^2 + 
{T_2 \over \sqrt{2} (2\pi)^4} \int_{M^{11}_D} C \wedge X_8. \label{bmbta}
\end{equation}
This topological action is not invariant under the gauge transformation $C
\rightarrow C + d \Lambda$. Brax and Mourad then introduced a boundary action
in the spirit of their worldvolume construction for the open membrane action as
\begin{equation}
\Delta S_T = - {2 \sqrt{2} \over  \kappa^2} 
\int_{\partial M^{11}_D} B\wedge G^2 +
{T_2 \over \sqrt{2} (2\pi)^4 }\int_{\partial M^{11}_D} B\wedge X_8.
\end{equation}
Then the total action $S_T + \Delta S_T$ is indeed invariant under the above
gauge transformation supplemented with $B \rightarrow B - \Lambda$.

However, the boundary term $\Delta S_T$ is not invariant under the
variation of Eq.\ (\ref{bmv}). Gauge and gravitational Anomalies arise.
They are on the ($x^{11} = 0$) component  of the boundary
\begin{equation}
\delta \Delta S_T = - {4\over 6 \kappa^2} \bigg({1 \over 8\pi T_2}\bigg)^3
\int_{M^{10}} Q^1_2 \wedge {\hat{I}^2_4  \over 4}
+ {1 \over 48 (2\pi)^5} \int_{M^{10}} Q^1_2 \wedge X_8.\label{bmkeyone}
\end{equation}
The striking feature of the present approach is that the last term on the right
hand of the above equation is independent of any coupling constant. This will
determine which bulk action one should use in the ``downstairs" approach when
the anomaly-free condition of M-theory is imposed.

From section 2, we have the variation of quantum effective action given by
 Eq.\ (\ref{quantuma})
\begin{equation}
\delta \Gamma = - {1 \over 2} {1 \over 48 (2\pi)^5} \int_{M^{10}} Q^1_2 \wedge
\bigg(- {\hat{I}^2_4 \over 4} + X_8\bigg).
\end{equation}
We expect that $\delta \Delta S_T + \delta \Gamma$ vanishes. But this is 
impossible since the terms involving $X_8$ do not cancel each other 
exactly because the last term in Eq.\ (\ref{bmkeyone}) is too large by a factor
2\footnote{It is clear  from their paper that Brax and Mourad overcounted
$\delta \Gamma$ by a factor 2 and forced themselves 
to agree with Ho\u{r}ava and Witten's bulk ``downstairs" action. Otherwise,
they would obtain Conrad's action much earlier.}. This indicates that 
Ho\u{r}ava and Witten's ``downstairs" bulk action is too large by a factor 2. 
Therefore, Conrad's ``downstairs" bulk action is correct. Using Conrad's
``downstairs" bulk action, we have $\delta \Delta S_T + \delta \Gamma = 0$
provided $T_2 = \bigg({(2\pi)^2 \over 2 \kappa^2}\bigg)^{1/3}$, i.e., the M-2
brane tension for $m = 1$ obtained in the previous section. With this result,
one can examine that the present G-value on the boundary is the same as that
obtained in the previous section for $\alpha = 1, m = 1$ and $n = 1$. 

The other consistency check showing the correctness of Ho\u{r}ava
and Witten's ``upstairs" bulk action is to go to the weak coupling limit of the
heterotic string. Then we expect to obtain the low energy effective action of
the $E_8 \times E_8$ heterotic string.  
 The ``upstairs" low energy effective 
action of M-theory on $R^{10} \times S^1/Z_2$ is
\begin{equation}
S = - {1\over 2 \kappa^2} \int_{M^{11}_U} d^{11} x \sqrt{-g_M} R - {1 \over 4
\lambda^2} \sum_i \int_{M^{10}_i} {\rm tr} F^2_i + \cdots. \label{usa}
\end{equation}
  
In the weak coupling limit, M-theory metric $d S^2_M = g_{M mn} d x^m dx^n$
can be expressed in terms of the heterotic metric as
$d S^2_M = g^{4/3}_H (dx^{11})^2 + g^{- 2/3}_H g_{\mu\nu} d x^\mu dx^\nu$ with
$g_H$ the coupling constant of the heterotic string. If we denote the radius of
the circular 11-th dimension as $R_0$, the physics radius in terms of M-theory
metric is $R_{11} = g^{2/3}_H  R_0$ from the above metric relation. As in
\cite{deatwo}, $R_0$ is conventionally chosen as $\sqrt{\alpha'}$ with $\alpha'$
the string constant. With this choice and other well-established relations
given in \cite{deatwo}, 
the eleven dimensional gravitational constant $\kappa$, M-2 and M-5 brane 
tensions can all be expressed in terms of one single string constant $\alpha'$ 
(for detail about these derivations, see \cite{deatwo}). Then, we have, with the
$Z_2$ symmetry imposed on all fields during the dimensional reduction,   
\begin{equation}
S = - {1\over (2\pi)^7 \alpha'^4} \int d^{10} x \sqrt {-g} e^{- 2\phi} 
\bigg[ R + {\alpha'\over 8} {\rm tr}F^2 + \cdots\bigg]\label{oureta},
\end{equation}
where the present $\eta = (4\pi)^5$ is used. 
 From the above action, we have the ten dimensional 
gravitational constant $\kappa_{10}$ satisfying $2 \kappa_{10}^2 = (2\pi)^7
\alpha'^4 g^2_H$. This is indeed the correct relation found independently in
ten dimensions, lending firm support to our claim.

The relative factor $\alpha' / 8$ in front of the kinetic term of
gauge fields in the above action now answers once for all that  one Yang-Mills 
instanton corresponds to {\it one}  Strominger's heterotic fivebrane
\cite{str} rather than eight ones. Many physicists once argued that this factor
should be\footnote{To my knowledge, this was pointed out first by  
Paul Townsend. This relative factor $\alpha'/8$ was previously given
explicitly, following the work of \cite{dixdp,tow},
in \cite{dufkl} in the action which is essentially identical to 
Eq.\ (\ref{oureta}).} $\alpha' /8$ rather than $\alpha'$ as used in \cite{str}.
However, if Ho\u{r}ava and Witten's $\eta = 2^7 \pi^5$ is used in the above 
as in \cite{deatwo},  a factor $\alpha'/4$ will be obtained instead. 

It is now clear that if we take the ten dimensional action Eq.\ (\ref{oureta})
as a standard one, one would be forced to choose $R_0 = \sqrt{\alpha'}$ if we
insist that the M-theory action \ (\ref{usa}) be reduced to the action 
\ (\ref{oureta}) in the weak coupling limit.   
    
\section{Acknowledgements}
I am grateful to M. J. Duff for discussions, 
for suggesting me some consistency checks and for reading the manuscript.
I would also like to thank K. Benakli for reading the manuscript.

\end{document}